\documentclass{epl}
\usepackage{epsfig,amsmath,amssymb,array,dcolumn,subfigure,rotating}

\def\brho{{\mbox{\boldmath $\rho $}}}

\def\bnabla{{\mbox{\boldmath $\nabla $}}}
\def\ul#1#2{\textstyle{\frac#1#2}}
\newcommand {\vct}[1] {\mathbf {#1}}

\shorttitle{Charge disorder and Coulomb interactions}
\title{Electrostatic disorder-induced interactions  in inhomogeneous dielectrics}

\author{ Rudolf Podgornik \inst{1,2} and Ali Naji \inst{3}}
\institute{
\inst{1} Dept. of Physics, Faculty of Mathematics and Physics, \\
University of Ljubljana,  SI-1000 Ljubljana,  Slovenia\\
\inst{2} Dept.  of Theoretical Physics, 
J. Stefan Institute, SI-1000 Ljubljana, Slovenia \\
\inst{3} Dept. of Chemistry and Biochemistry,  University of California, Santa Barbara, \\
CA 93106-9510, USA }

\begin{document}

\maketitle
\vspace{-0.5cm}
\begin{abstract}
We investigate the effect of quenched surface charge disorder on electrostatic 
interactions between two charged surfaces in the presence of dielectric inhomogeneities and added salt. 
We show that  in the linear weak-coupling regime (i.e., by including mean-field and Gaussian-fluctuations
contributions),  the image-charge effects lead to a non-zero disorder-induced 
interaction free energy between two surfaces of equal mean charge that
 can be repulsive or attractive depending on the dielectric mismatch across the
bounding surfaces and the exact location of the disordered charge distribution.
\end{abstract}
\vspace{-0.5cm}

Electrostatic interactions are one of the two fundamental components  of the DLVO theory of colloidal stability 
\cite{VO,LesHouches}. They are standardly described by the Poisson-Boltzmann (PB) theory \cite{VO,andelman95}
embodying the {\em mean-field} approach to classical charged systems. Mean-field interactions between like-charged
macroions are repulsive in nature and thereby tend to stabilize solutions of charged macroions.  In strongly coupled systems
(e.g., when multivalent counterions are present),  electrostatic interactions  however induce strong attractive forces 
between like-charged macroions 
\cite{ali-review,hoda-review}, and thus act more like Lifshitz-van
der Waals interactions that tend to destabilize charged solutions.  This attraction can not be captured by
the mean-field approach and a new paradigm dubbed the {\em strong-coupling limit} \cite{netz-SC,grosberg} was introduced 
to describe the equilibrium properties of Coulomb fluids when the mobile counterion charges become large. 
The crossover from the mean-field  Poisson-Boltzmann description to the strong-coupling limit
is governed by a single dimensionless electrostatic coupling parameter, which is given by 
the ratio of the Bjerrum length
(identifying Coulombic  interaction between ions themselves) and the Gouy-Chapman length (describing
electrostatic interaction between ions  and the charged macroion surface) \cite{ali-review}. 
Electrostatic interactions between charged macroions in the mean-field and the strong-coupling limit thus unfold into a much richer
structure than conveyed for many  years by the DLVO paradigm. The collapse of a highly charged polyelectrolyte, such as DNA, in 
the presence of multivalent counterions is the most dramatic example of unexpected and counter-intuitive features of 
the strong-coupling electrostatics  \cite{Bloom,grosberg}.

Recently we added a new twist to the theory of electrostatic interactions in charged systems \cite{ali-rudi}: not only can 
electrostatic interactions between like-charged macroions turn from repulsive to attractive due to strong-coupling 
counterion-induced correlations, but we showed that in the
case of no added salt and no image interactions, 
the {\em quenched disordered} distribution of surface charges on the macroions can induce an  
additive attractive interaction in the strong-coupling limit even if the
mean charge of the macroions is {\em zero}. This effect is due to 
the nonlinear features of the average over quenched disorder of
the distribution of charges on the macroion surfaces.  
Such quenched distributions of macroion charge have been invoked 
recently  in experimental investigations of interactions between solid surfaces in the presence of charged surfactants 
\cite{manne,klein}. The patterning of interacting surfaces by quaternary ammonium surfactants in these experiments is
highly disordered, depends on the method of preparation and has basic implications also for the forces that act between other
types of hydrophilic surfaces with mixed charges, most notably in biological as well as  in
synthetic systems. Motivated by these observations we will now try to develop the theory of electrostatic interactions
in systems with quenched disordered macroion charge distributions further.  

Here we will consider the effects of {\em
added salt} and of {\em image interactions} due to dielectric inhomogeneities on the disorder-induced interaction
between two charged walls of equal mean charge density.
We shall focus only on the weak-coupling regime  and evaluate interaction free energies
up to the first-loop (Gaussian-fluctuations) contribution around the linearized 
mean-field (Debye-H\"uckel) solution.  
We will show that in general image interactions have
a pronounced effect on the way disordered charge distributions bring about electrostatic interactions in salt solution,
their most notable effect being that they can induce non-monotonic interactions as a function of the spacing between the
interacting surfaces. These results markedly contrast the weak-coupling results obtained in the
absence of added salt and image-charge effects \cite{ali-rudi}, where the disorder contribution turns out to be nil.

\begin{figure}[ht]
\onefigure[width=6cm]{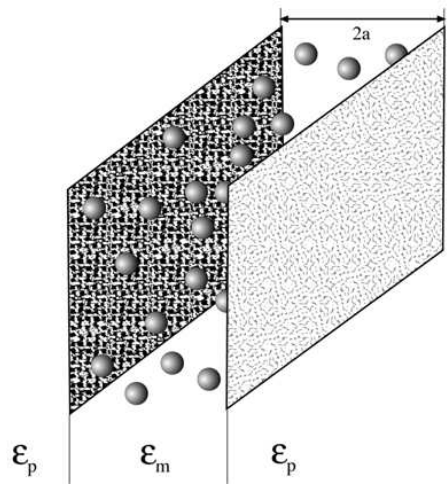}
\caption{Geometry of a system composed of two surfaces with disordered charge distribution at separation $D=2a$ 
with different dielectric constants for the interior region (where salt ions are present), $\varepsilon_m$,  and  for
the  exterior region, $\varepsilon_p$. }
\label{fig0}
\end{figure}

Assume a monovalent salt solution confined between two charged surfaces at $z=\pm a$.  Assume furthermore that the
surface charges are small such that the linearization approximation is valid.  The corresponding grand canonical
partition function
in the field of an external fixed charge distribution, $\rho(\vct r)$, is given in the functional integral representation as
\begin{equation}
{\mathcal Z} = \int {\mathcal D}[\phi(\vct r)] ~e^{- {\mathcal S}[\phi(\vct r)]},
\end{equation}
where $\phi(\vct r)$ is a fluctuating potential field and the linearized effective Hamiltonian reads \cite{action}
\begin{equation}
{\mathcal S}[\phi(\vct r)] \cong \frac{\beta \varepsilon_{0}}{2} \int d \vct
r~ \varepsilon({\mathbf  r})\bigg[ (\bnabla \phi(\vct r))^{2} + \kappa^{2}({\mathbf  r}) \phi^{2}(\vct r)\bigg]  + i
\beta \int d \vct r~ \rho(\vct r) \phi(\vct r).
\label{act-1}
\end{equation}
Here we take the dielectric constant $\varepsilon = \varepsilon({\mathbf  r})$ to have a value, $ \varepsilon_{p}$, 
 in the two semi-infinite regions (i.e. $|z|>a$), which is in general 
 different from, $ \varepsilon_{m}$, assumed in between the 
charged surfaces. Accordingly,  the inverse Debye screening length is 
$\kappa({\mathbf  r}) = \kappa = \sqrt{8 \pi \ell_{\mathrm{B}} n}$ in between the two surfaces 
(with $\ell_{\mathrm{B}}=e^2/(4\pi \varepsilon_0 \varepsilon k_{\mathrm{B}}T)$ being  the 
Bjerrum length  
and $n$ the salt concentration), and zero otherwise. We
assume that the surface charge distribution, $\rho(\vct r)$,  has a quenched disordered component. 
The average over quenched disorder is done as
\begin{equation}
{\mathcal F} = - k_{\mathrm{B}}T ~\overline{\ln{{\mathcal Z}}}, 
\end{equation}
where the disorder average is defined here {\em  via} a Gaussian probability distribution as
\begin{equation}
\overline{(\dots)} = \int {\mathcal D}[\rho(\vct r)] (\dots) e^{- \ul12 \int
d \vct r~g^{-1}(\vct r) \left( \rho(\vct r) - \rho_{0}({\mathbf  r})\right)^{2}}. 
\end{equation}
Note that  $\rho_{0}({\mathbf  r})$ represents  the mean charge density and $g(\vct r)$
gives the  disorder variance around the mean value.
Since all the functional integrals in the expression  for the disorder-averaged free energy are Gaussian, the free
energy follows straightforwardly as
\begin{equation}
{\mathcal F} = - k_{\mathrm{B}}T ~\overline{\ln{{\mathcal Z}}} = {\ul\beta2} ~{\rm Tr}~{g(\vct
r) {\mathcal G}(\vct r, \vct r')} + {\textstyle\frac{k_{\mathrm{B}}T}2}~{\rm Tr}\ln{{\mathcal G}^{-1}(\vct r, \vct r')} +
{\ul12} \! \int\!\!\!\int \!\! d \vct r \,d \vct r'
\rho_{0}(\vct r) {\mathcal G}(\vct r, \vct r') \rho_{0}(\vct r').
\label{trace-1}
\end{equation}
Here we have defined the inverse of the operator $\beta\varepsilon\varepsilon_{0}(-
\nabla^{2} + \kappa^{2})$ as the Green function that satisfies 
\begin{equation}
\beta\varepsilon_{m}\varepsilon_{0}(- \nabla^{2} + \kappa^{2}) {\mathcal G}(\vct r, \vct r') =
\delta^{3}(\vct r - \vct r')
\end{equation}
with the appropriate  boundary conditions of the continuity of derivatives
multiplied by the dielectric constants at the surfaces with dielectric
discontinuity. The disorder-averaged partition function could also be obtained through the replica formalism
\cite{ali-rudi} but the direct integration approach is much more straightforward in the case of linearized
effective Hamiltonian, Eq. (\ref{act-1}). In the second and third terms of Eq. \ref{trace-1}, we recognize the usual 
fluctuational and linearized mean-field Debye-H\"uckel (DH) contributions respectively. The first term is thus
stemming from the effects of the disorder. Let us evaluate it explicitly and anlayze its consequences. 

Because of transverse isotropy, the following Fourier decomposition for the Green function is valid
\begin{equation}
{\mathcal G}(\vct r, \vct r') = \int \frac{d^{2}  Q}{(2\pi)^{2}} \, {\mathcal G}(\vct Q; z, z')
~e^{- \imath \vct Q \cdot (\brho - \brho')},
\end{equation}
with $z$ and $z'$ denoting the normal coordinate to the surfaces and $\brho = (x, y)$, the transverse coordinates.  We now
evaluate two Green functions corresponding to the cases
\begin{enumerate}
\item when the disordered charge distribution is contained
within the medium $\varepsilon_{m}$, and 
\item when the disordered charge distribution is contained
within the medium $ \varepsilon_{p}$.
\end{enumerate} 
These Green functions can be derived straightforwardly by using the methods described in Ref. \cite{green} as
\begin{equation}
{\mathcal G}(\vct Q; z, z') = \frac{1}{2 \beta \varepsilon_{m}\varepsilon_{0}~ u} \left[ e^{- u
\vert 
z - z'\vert} + \frac{2 \alpha e^{- 4 u a}}{1 - \alpha^{2} e^{- 4 u a}} \left( 
e^{2 u a} \cosh{u(z + z')} + \alpha \cosh{u (z - z')}\right)
\right]
\label{green-1}
\end{equation}
for the case (i), and as 
\begin{equation}
{\mathcal G}'(\vct Q; z, z') = \frac{1}{2 \beta \varepsilon_{p}\varepsilon_{0}~ Q} \left[ e^{- Q
\vert z - z'\vert} - \frac{\alpha (1 - e^{- 4 u a})}{1 - \alpha^{2} e^{- 4 u a}} \left( e^{-Q \vert a + z'\vert + Q(a+z)}\right)
\right],
\label{green-2}
\end{equation}
for the case (ii). Here
\begin{equation}
\alpha(Q) = \frac{\varepsilon_{m} u(Q) - \varepsilon_{p} Q}{\varepsilon_{m} u(Q) + \varepsilon_{p} Q} \qquad {\rm
with} \qquad u^{2}(Q) = Q^{2} + \kappa^{2}.
\end{equation}
Furthermore, we assume that the disorder variance is surface distributed and thus
\begin{equation}
g(\vct r) = G ~\delta(z + a) + G~ \delta(z - a).
\end{equation}
It can  reside either inside the slab (e.g., at $a - \delta$) or outside the slab 
(e.g., at $a + \delta$ for arbitrarily  small $\delta$).
The only difference in the calculation is whether expressions  (\ref{green-1}) or  (\ref{green-2}) are used when
evaluating the  first trace in Eq. \ref{trace-1}.  Subsequently, one obtains the disorder-induced part of the free
energy for the two aforementioned cases of the location of the disordered charge distribution as 
\begin{equation}
{\ul\beta2} {\rm Tr}~{g(\vct r) {\mathcal G}(\vct r, \vct r')} = \frac{
{G} S}{4\pi~\varepsilon_{m}\varepsilon_{0}} \int_{0}^{\infty} {Q dQ}\, \frac{\alpha~(1 + \alpha)^{2}
~e^{- 4 u a}}{u (1 - \alpha^{2} e^{- 4 u a})} = \frac{
{G} S~\kappa}{4\pi~\varepsilon_{m}\varepsilon_{0}}~F_{\mathrm{(i)}}(\kappa a),
\label{result-1}
\end{equation}
for the case (i),  and as 
\begin{equation}
{\ul\beta2} {\rm Tr}~{g(\vct r) {\mathcal G}'(\vct r, \vct r')} = \frac{
{G} S}{4\pi~\varepsilon_{p}\varepsilon_{0}} \int_{0}^{\infty}{Q dQ}\, \frac{\alpha~(1 - \alpha^{2})
~e^{- 4 u a}}{Q (1 - \alpha^{2} e^{- 4 u a})} = \frac{
{G} S~\kappa}{4\pi~\varepsilon_{p}\varepsilon_{0}}~F_{\mathrm{(ii)}}(\kappa a),
\label{result-2}
\end{equation}
for the case (ii), where $S$ is the total area of the two bounding surfaces. In both of the above expressions we have subtracted the
part of the free energy that does not depend on the separation $a$ since we are only interested in the interaction
free energy. 

It is thus immediately obvious that some asymmetry should
exist in the system (either different dielectric constants in between and outside the surfaces, or salt in between and
no salt outside, etc.) in order that the disorder contribution to the free energy becomes non-zero. Also obviously in both cases
(i) and (ii) if $\alpha = 0$, there is no disorder-induced interaction. If $\alpha = -1$, then in the case (i), the interaction is
zero, but not in the case (ii).

\begin{figure}[ht]
\onefigure[width=14cm]{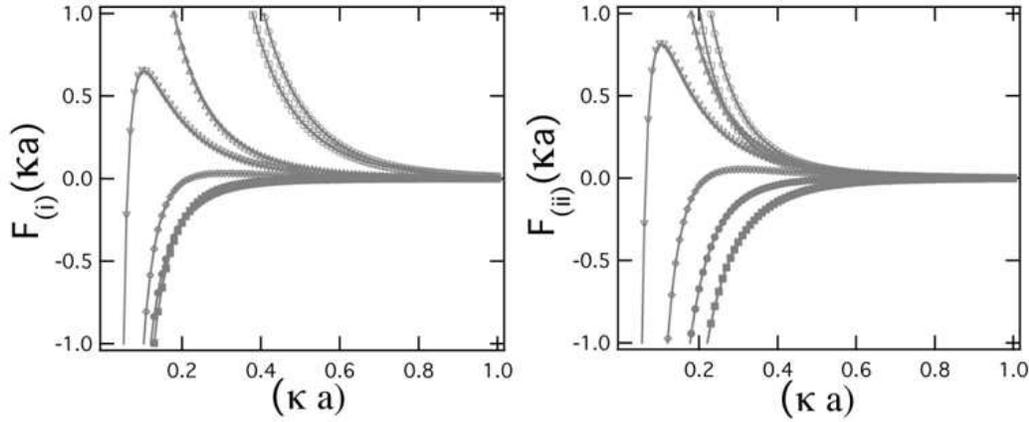}
\caption{Numerical evaluation of the disorder-induced interaction between two charged surfaces, $F_{\mathrm{(i)}}(\kappa a)$ 
and $F_{\mathrm{(ii)}}(\kappa a)$ from Eqs. \ref{result-1} (left) and \ref{result-2} (right) for $0 < \kappa a < 1$. The values
of the ratio $\varepsilon_{m}/\varepsilon_{p}$ are $0.2, 0.4, 0.6, 0.8, 1.0, 5.0$ and 10.0
 (from top to bottom). The non-monotonic character
of the disorder-induced interaction is clearly discernible. Its details depend crucially on the ratio $\varepsilon_{m}/\varepsilon_{p}$.}
\label{fig1}
\end{figure}

The dependence of the disorder part of the free energy, that is Eqs. \ref{result-1} and \ref{result-2}, 
on the dimensionless separation between the surfaces, $\kappa
a$, is shown in Fig. \ref{fig1}. The most interesting feature of the disorder-induced interaction free energy 
is that the interaction can be non-monotonic and that it depends critically on the
ratio of the two dielectric constants. For small and large values of $\kappa a$, the disorder interaction free energy
assumes simple limiting forms as we show later.

Putting the above results together with the fluctuational and mean-field contributions, where we assume that the  surface charge
distribution has a mean value given by
\begin{equation}
\rho_{0}(\vct r) = \sigma ~\delta(z - a) + \sigma~ \delta(z + a),
\end{equation} 
we obtain the following expressions for the interaction free energy, Eq. \ref{trace-1},  in the case (i) and (ii) respectively, i.e.
\begin{equation}
\frac{\mathcal F}{S} = \frac{
{G}}{4\pi\varepsilon_m \varepsilon_{0}} \int_{0}^{\infty}\! \frac{{Q dQ}~\alpha~(1 + \alpha)^{2}
~e^{- 4 u a}}{u (1 - \alpha^{2} e^{- 4 u a})} + \frac{k_{\mathrm{B}}T}{4\pi} \int_{0}^{\infty}\!\!\!\!\!{Q dQ} \ln{(1 - \alpha^{2}
~e^{- 4 u a})} + \frac{\sigma^{2} }{
\varepsilon_m\varepsilon_{0} \kappa} (\coth{\kappa a} - 1).
\label{res-1}
\end{equation}
and
\begin{equation}
\frac{\mathcal F'}{S} = \frac{
{G}}{4\pi\varepsilon_{p}\varepsilon_{0}} \int_{0}^{\infty}\!  \frac{Q dQ~\alpha~(1 - \alpha^{2})
~e^{- 4 u a}}{Q (1 - \alpha^{2} e^{- 4 u a})} + \frac{k_{\mathrm{B}}T}{4\pi} \int_{0}^{\infty}\!\!\!\!\!{Q dQ} \ln{(1 - \alpha^{2}
~e^{- 4 u a})} + \frac{\sigma^{2} }{
\varepsilon_m\varepsilon_{0} \kappa} (\coth{\kappa a} - 1).
\label{res-2}
\end{equation}
These are the final results of our calculation. We note here that the only
approximation involved in the derivation of the above results is the linearization approximation in the Coulomb field
action, Eq. \ref{act-1}, that makes them valid only in the weak-coupling limit,   i.e.,  for small
mean surface charge density $\sigma$ and low counterion valency.

Let us assess the importance of the disorder-induced interaction by considering a few illuminating limiting cases, the
general form being given numerically in Fig. \ref{fig1}.
In the case of vanishing salt or small separations, $\kappa a \rightarrow 0$, one gets in the case (i), where 
disorder is located inside the slab of dielectric constant $\varepsilon_{m}$,  
\begin{equation}
{\ul\beta{2}} ~{\rm Tr}~{g(\vct
r) {\mathcal G}(\vct r, \vct r')} = \frac{G ~S ~\varepsilon_{m} (\varepsilon_{m} - \varepsilon_{p})}{4 \pi \varepsilon_{0}
(\varepsilon_{m} +\varepsilon_{p})^{3} ~a}~f\!\left(\frac{\varepsilon_{m}  - \varepsilon_{p} }{\varepsilon_{m} +
\varepsilon_{p}} \right),
\label{res-12}
\end{equation}
where
\begin{equation}
f(\alpha) \equiv \int_{0}^{\infty} \frac{du~e^{-u}}{(1 - \alpha^{2} e^{-u})}.
\end{equation}
The above limiting form is valid only if $\varepsilon_{m} \neq \varepsilon_{p}$. Otherwise higher order terms come into play.
The second case (with the disorder located outside the slab and in the medium $\varepsilon_{p}$) 
leads to exactly the same free energy and thus in this limit,  there is no difference in the disorder-induced
 interaction whether the disordered charge distribution is within medium $\varepsilon_{m}$ or $\varepsilon_{p}$.
Obviously in this limit, the disorder-induced part of the interaction falls  off  inversely with the first power of the
separation, $D=2 a$, to be compared with the inverse-square decay in the case of the zero-frequency van-der-Waals 
(fluctuational) term.
Its sign depends on the values of both dielectric constants. One should also
note that in this limit,  the disorder and the mean-field term combine, yielding
\begin{equation}
\frac{\mathcal F}{S} \simeq 
 \frac{\sigma^{2} }{
\varepsilon_{m}\varepsilon_{0} \kappa^{2} a} \left( 1 + \frac{G ~\varepsilon_{m}^{2} (\varepsilon_{m} -
\varepsilon_{p}) \kappa^{2}}{4 \pi \sigma^{2} (\varepsilon_{m} +\varepsilon_{p})^{3} }~f\!
\left(\frac{\varepsilon_{m} - \varepsilon_{p} }{\varepsilon_m +
\varepsilon_{p}} \right)\right).
\label{futu-1}
\end{equation}
It would thus seem that the disorder merely renormalizes the square of the charge density. But since the disorder term
can be either positive or negative, depending on the value of $\varepsilon_{m}  - \varepsilon_{p}$, one can not claim
that the only effect of disorder in this limit is the disorder-renormalization of the
mean surface charge, since the whole expression Eq. \ref{futu-1} can not be written as
proportional to $(\sigma_{R})^{2} = (\sigma - \sigma')^{2}$, which is by definition always positive, where $\sigma'$
would indicate the disorder dependent terms. The disorder in this limit therefore does not simply renormalize the
surface charge and can lead to attractive or repulsive interactions, depending on the sign of $\varepsilon_{m} -
\varepsilon_{p}$.
However, an important consequence of the disorder effects  in this limit is that it induces interactions even between
nominally uncharged surfaces with a mean charge  density $\sigma = 0$. These interactions have the same dependence on the
separation as the mean-field DH term in this limit, except that they can be either repulsive or attractive 
depending again on the difference $\varepsilon_{m}  - \varepsilon_{p}$. 
Nominally neutral surfaces thus exhibit electrostatic-like  interactions induced {\em solely} by the 
variance of the charge distribution, not its mean value!

In the opposite limit of large salt or large separations, $\kappa a \rightarrow \infty$, one remains with
\begin{equation}
\frac{\mathcal F}{S} = \frac{G ~e^{-4 \kappa a}}{\pi \varepsilon_{0} \varepsilon_{m} ~4 a} -
\frac{k_{\mathrm{B}}T~\kappa^{2}}{16 \pi ~(\kappa a)}~e^{- 4 \kappa a} + 2 \frac{\sigma^{2}}{\varepsilon_m\varepsilon_{0} \kappa}~e^{- 2 \kappa a}.
\label{res-11}
\end{equation}
for the  case (i).  The disorder-induced component (first term) has the same separation dependance as the
standard screened zero-frequency van-der-Waals (vdW) interaction (second term), but is shorter 
ranged than the corresponding mean-field DH term (third term).  
Also in this limit, the disorder-induced interaction is always {\em repulsive}, which means that the overall interaction
can change sign upon increase
of the separation, as is already apparent from Fig. \ref{fig1}. The interesting point now is that the disorder-induced
term clearly renormalizes the fluctuational (van-der-Waals)  contribution, since it has  the same separation-dependance 
as the zero-frequency van-der-Waals term but with the opposite sign.  

For the second case (ii) and in the same limit of $\kappa a \rightarrow \infty$, we obtain
\begin{equation}
\frac{\mathcal F'}{S} =   \frac{G}{\pi \varepsilon_{0} \varepsilon_{m}}~\sqrt{\frac{\pi}{8\kappa a }}~e^{- 4  \kappa a}
- \frac{k_{\mathrm{B}}T~\kappa^{2}}{16 \pi ~(\kappa a)}~e^{- 4 \kappa a} + 2 \frac{\sigma^{2}}{\varepsilon_m\varepsilon_{0} \kappa}~e^{- 2 \kappa a}.
\label{res-22}
\end{equation}
Again the disorder part of the interaction (first term) 
has almost the same functional dependence on the intersurface separation as
the zero-frequency van-der-Waals part (second term). 

One can thus make a general conclusion that in the limit $\kappa a \rightarrow 0$,
the disorder-induced component of the interaction free energy effectively behaves like the mean-field contribution,
while in the limit $\kappa a \rightarrow \infty$, it behaves like the fluctuational (vdW) contribution. 
In a certain sense, the disorder-induced interaction thus {\em interpolates} between mean-field and
fluctuational interactions. All this is of course valid only in the weak-coupling limit and one can not apply these
conclusions to the disordered strong-coupling regime \cite{ali-rudi}. 
The analysis of the interplay between disorder-induced effects and 
image-charge effects in the strong-coupling limit will be left for a separate exercise. 

The results derived above, apart from the effects due to finite salt concentration in 
between the apposed charged surfaces, 
clearly differ from those obtained in the absence of dielectric inhomogeneities \cite{ali-rudi}, where the 
mean-field contribution was shown to be decoupled from the disorder effects. The conclusion reached in
Ref. \cite{ali-rudi} is thus limited to disordered charge distributions immersed in a single dielectric medium
without any inhomogeneities in the corresponding static dielectric constant.

Note that if the mean surface charge is zero ($\sigma=0$), 
then according to Eqs. \ref{res-11} and \ref{res-22}, the equilibrium spacing between
surfaces is given by the competing disorder-induced and fluctuational interaction parts for large $\kappa a$. 
In case (i), Eq. \ref{res-11} obviously implies no finite equilibrium
spacing. The interaction is monotonic and its sign depends on whether the ratio 
$4 G/(\kappa \varepsilon_{m}\varepsilon_{0} k_{\mathrm{B}}T)$ is 
bigger or smaller than one. In  case (ii), Eq. \ref{res-22}, an optimal surface separation exists since the disorder and fluctuational
contributions do not have exctly the same separation-dependence.

In all the limiting cases addressed above, 
the disorder-induced part of the interaction can be masked by either mean-field or fluctuational
terms in the total interaction, which would make its effects particularly difficult to pinpoint experimentally. Its most
important feature, though, is the non-monotonic character of the interaction at intermediate separations (see Fig. \ref{fig1}).  
In the case of interactions between charged interfaces, this feature may be important for the stability of
planar charged macromolecular assemblies such as lipid bilayers.

\acknowledgments 
One of the authors, RP, would like to acknowledge an illuminating discussion with Henri Orland that eventually lead to
this work. RP would like to acknowlegde the financial support of the Agency for Research and Development of Slovenia
under the grant P1-0055 (C). A.N. gratefully acknowledges generous support provided by the 
J. Stefan Institute, Ljubljana, for his visit to the Institute, as well as research 
funds from the DPG German-French Network.  


\end{document}